# Self-healing and transformation characteristics of obstructed Hermite-Gaussian modes composed structured beams

Suyi Zhao[a,b,c], Zilong Zhang[a,b,c,*], Yuan Gao[a,b,c], Xin Wang[a,b,c], Yuchen Jie[a,b,c], and Changming Zhao[a,b,c]

[a] School of Optics and Photonics, Beijing Institute of Technology, 5 South Zhongguancun Street, 100081 Beijing, China
[b] Key Laboratory of Photoelectronic Imaging Technology and System, Ministry of Education of People's Republic of China
[c] Key Laboratory of Photonics Information Technology, Ministry of Industry and Information Technology

*Corresponding author: zlzhang@bit.edu.cn*

The self-healing property of laser beams is of great interest. And a laser beam with spatial structures is also widely concerned due to its important applications in lots of areas. We theoretically and experimentally investigate the self-healing and transformation characteristics of obstructed structured beams composed by incoherent or coherent superposition of multiple Hermite-Gaussian (HG) modes. It's found that partially obstructed single HG mode can recover itself or transfer to a lower order in the far-field. When the obstacle retains one pair of edged bright spots of HG mode in each direction of its two symmetry axes, the beam structure information (number of knot lines) along each axis can be restored. Otherwise, it will be transferred to the corresponding low-order HG mode or multi interference fringes in the far-field, according to the interval of the two most edged remaining spots on each axis. It's proved that the above effect is induced by the diffraction and interference results of the partially retained light field. This principle is also applicable to multi-eigenmodes composed beams with special customized optical structures. Then the self-healing and transformation characteristics of the partially obstructed HG modes composed structured beams are investigated. It's found that the HG modes incoherently composed structured beams have a strong ability to recover themselves in the far-field after occlusion. These investigations can expand the applications of optical lattice structures of laser communication, atom optical capture, and optical imaging.

## 1. Introduction

The self-healing characteristic of a laser beam is that the beam can reconstruct itself when it encounters obstacles in the process of propagation. Self-healing properties have important applications in laser communication [1], quantum entanglement propagation [2], micromanipulation [3], and optical imaging [4]. This property is firstly found with non-diffracting beams, such as Bessel beam [5-7], Airy beam [1,8-10], dark and anti-dark diffraction-free beams [11,12]. Apart from being obstructed by opaque obstacles, non-diffracting beams also show robustness in the strong focusing medium [13], scattering medium [14], and turbulent environment [15,16]. In order to obtain symmetrical mode structures and uniform light intensity distribution to improve the application value, some non-diffracting beam arrays [17] with good self-healing characteristics were proposed. Taking the Gaussian beam as an example, some general conclusions about beam self-recovery are obtained by establishing the mathematical model of orthogonal components [18] or from the view of wave optics [19]. At the same time, due to the propagation invariance of eigenmode laser beams, the beam shapes of high-order HG beam [20,25] and LG beam [21-24] have significant resilience to disturbances introduced by obstructing objects. In addition, vortex beams with special phase singularity and orbital angular momentum [27-29] have been proved to have the ability of self-recovery. Previous studies mainly focused on fully coherent laser beams, and then the concept of self-reconstruction was extended to partially coherent beams. As long as the coherent area in the beam cross-section is reduced to smaller than the area of the obstacle, its intensity distribution, degree of coherence, and state of polarization can still be reconstructed by its self-healing property [30-32]. Subsequently, the self-healing properties of partially coherent LG beam [33], partially coherent HG beam [34], and partially coherent beams of other correlation models [35] have also been proposed. In addition to single-mode beams, the self-healing properties of optical lattice structures formed by the coherent superposition of multiple modes have also been studied. Optical ring lattice as a structured beam composed of LG modes has self-healing property, which was firstly proposed by Pravin Vaity [36]. Litvin predicted its self-healing distance and the dependence of the recovery position and the size of obstacles [37]. Subsequently, Zhao studied the evolution behavior of the structured beams generated by the superposition of high-order LG beams and discovered its quasi-nondiffracting property [38].

As propagation-invariant laser beams, the self-healing properties of single HG and LG beams have been studied extensively. Early studies roughly attributed the self-healing of the beam to diffraction in the process of propagation in free space [20] and explained the self-healing characteristics of the laser beam from the perspective of internal transverse power flow [37]. A class of studies explained the influence of obstacle's position on the beam cross-section through spatial filtering theory, that is the beam can achieve self-healing when the external high-frequency component is retained [21-23]. Another class of studies decomposed structured beams into multiple traveling waves to explain their self-healing characteristics [24-26]. However, the above theories are not only limited to the variable conditions such as the position and area of obstacles and the propagation distance of beams but also not intuitive or concrete enough. More importantly, the theories are only aimed at a certain class of beam, while beams under different categories having similar physical properties should have a unified theory to explain. For the eigenmodes composed structured beam, the research on self-healing is only limited to the coherent superposition of LG modes and does not analyze the internal mechanism of structural recovery or distortion, let alone the self-healing characteristics of HG modes superposed structures.

In this work, we investigate the self-healing and transformation characteristics of complex structured beams generated by the coherent and incoherent superposition of HG modes and analyze the intrinsic reasons for the self-healing and transformation properties. Firstly, through theoretical analysis and experimental verification, we point out that the reconstruction of the partially obstructed single-mode beam is a process of combining the diffraction and interference of the remaining light field. Through the diffraction effect, the partially obstructed laser beam can realize the transverse flow of energy to the shaded area, which is helpful to realize the energy recovery of the obstructed area in the far-field. However, the reconstruction of beam structure depends on the interference effect of the light field in the unobstructed area. The interference between the outer regions of the beam can better restore the original light field structural information. The far-field reconstruction of the beam whose one edge is obstructed only depends on the diffraction effect of the remaining light field and will lead to the loss of the original structural information. This property can be extended to the coherent and incoherent superposition of multiple HG (or LG) modes. Once this kind of structured beam is partially occluded by obstacles, most component modes in the beam will be recovered after a certain distance of propagation, as the beam can retain most of the outer edge spots under the original width. Though few modes in the structured beam may be obstructed with edged parts, resulting in the decrease of the orders of these modes, the reconstructed beam still has a strong tendency to maintain its original pattern information by the intensity complementary of the other component modes in the structured beam. So, it enhances the self-reconstruction ability of the beam or the occlusion robustness of the beam. Both numerically simulated and experimental results prove the above characteristics of the single laser mode and the HG modes composed structured beams.

## 2. Theoretical analysis and simulations

### 2.1. Self-healing and transformation of obstructed single mode

In order to analyze the self-healing and transformation properties of the HG modes composed structured beam after the obstruction, its decomposed single component is studied at first. A laser beam with complex transverse intensity distribution can be regarded as the coherent or incoherent superposition of multiple eigenmodes (HG or LG modes). For the incoherent superposition, the beam intensity has a linear superposition of the component modes both before and after the obstruction. Then we can deal with each mode separately first and then add them together. And for the coherent superposition, we need to treat the beam as a single mode, whose field expression equation is a summation of the field of eigenmodes with an additional phase item.

Assuming that the beam waist of the structured beam is located at z = 0, its single component HG mode experiencing free transmission to the plane of the obstacle (z=L) can be expressed as follows:

$$E_b(x,y,z=L) = A_{mn} \frac{\omega_0}{\omega(L)} H_m(\frac{\sqrt{2}x}{\omega(L)}) H_n(\frac{\sqrt{2}y}{\omega(L)}) \cdot exp[-\frac{ik}{2R(L)}(x^2+y^2)]exp[i(m+n+1)\phi(L)]. \quad (1)$$

Where $A_{mn}$ is the amplitude coefficient, $\omega(z) = \sqrt{\omega_0^2(1+z^2/z_R^2)}$, $R(z) = (z^2+z_R^2)/z$ is the radius of curvature of the wavefront, $\phi(z) = arctan(z/z_R)$, and $z_R$ is the Rayleigh length, and $H_m$ and $H_n$ are Hermite polynomials of order m and order n respectively. And for the coherent superposition of HG modes, which is in the state of transverse mode-locking (TML), the mode's light field could be expressed by the equation below,

$$E_{TML} = \sum_{m,n} a_{m,n} HG_{m,n}(\cdot)$$
$$\cdot exp(i\phi_{m,n} + ikz + ik\frac{x^2+y^2}{R(z)} - i(m+n+1)\phi(z)), \quad (2)$$

where $HG_{m,n}(\cdot)$ is the pure intensity item of HG modes, $a_{m,n}$ is the coefficient of each mode, $\phi_{m,n}$ is the locking phase of each HG mode.

The light field passing through the obstacle can be expressed as below,

$$E_0(x,y) = E_b(x,y) \cdot T(x,y), \quad (3)$$

where $T(x,y)$ is the transmittance coefficient of the obstacle. The transformation matrix from obstacle plane to observation plane is as follows,

$$M = \begin{bmatrix} A & B \\ C & D \end{bmatrix} = \begin{bmatrix} 1-Z/f & s+Z-s\cdot Z/f \\ -1/f & 1-s/f \end{bmatrix}. \quad (4)$$

Here, $f$ is the focal length of a positive lens, which is used to obtain a far-field beam pattern, $s$ is the interval between the lens and the obstacle, and $Z$ is the distance from the lens to the observation plane. According to the Collins diffraction integral formula [39], the field distribution of the observation plane is as follows,

$$E_r(x_1, y_1, z_1) =$$
$$\frac{ik}{2\pi B} exp(-ikz) \times \iint E_0(x,y)$$
$$\cdot exp\left\{-\frac{ik}{2B}\left[A(x^2+y^2) - 2(xx_1+yy_1) + D(x_1^2+y_1^2)\right]\right\}dxdy. \quad (5)$$

The incoherent light intensity distribution of the observation plane can be obtained by directly superimposing the light intensity of each component mode,

$$I_r = \sum E_r(x_1, y_1) * E_r^*(x_1, y_1). \quad (6)$$

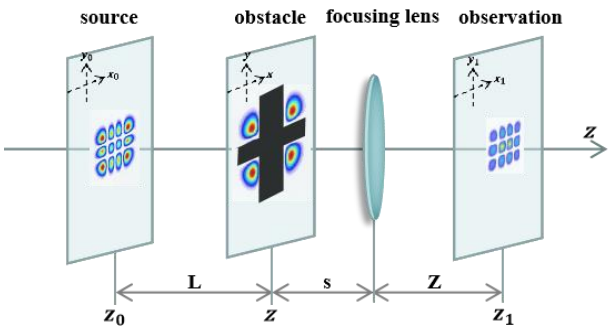

**Fig. 1.** Schematic diagram of the obstacle-lens propagation process of the structured beam.

The propagation progress process with obstacles and a focusing lens of structured beams is shown in Fig.1. In the numerical calculation example, the laser wavelength is 1064nm, the fundamental mode radius of the beam is 200μm, the interval between the beam waist and the obstacle is $L$=0.5m, the separation between the obstacle and the lens is $s$=0.02m, the propagation distance after the lens is $Z$=0.1221m, where is the exact position of the beam waist of the focused beam, and the focal length of the thin lens is $f$=0.1m. In the following research, the far-field light intensity distributions of the HG beam whose center or edge is blocked by obstacles with different widths are numerically calculated by using the equations above. In order to quantify the recovery degree of the beam, we calculated the similarity value [40], which is widely used in the study of self-healing, and defined it as:

$$S = \frac{\int_0^\infty \overline{E_r}(r,z) E_r^*(r,z) r dr}{\sqrt{\int_0^\infty \overline{E_r}(r,z)\overline{E_r^*}(r,z) r dr}\sqrt{\int_0^\infty E_r(r,z) E_r^*(r,z) r dr}}. \quad (7)$$

$\overline{E_r}(r,z)$ and $E_r(r,z)$ denote the optical fields of the beam with and without obstruction respectively, and $E^*$ denotes the complex conjugate of $E$.

The similarity analysis results between the far-field reconstructed beam of $HG_{40}$ obstructed by the center or the edge and the original beam are shown in Fig.2. It can be seen that the similarity between the reconstructed beam and the original beam will gradually decrease for both conditions with the increase of the width of the obstacles. In most cases, the similarity of the far-field reconstructed beams from the HG beams with edge occlusion is lower than that with center occlusion.

When the beam center is obstructed, the internal structure of the beam can be reconstructed in the far-field through diffraction and interference, but the relative intensity distribution of the reconstructed light spots has changed compared with the initial state, resulting in the decrease of similarity (as shown in Fig.2 (a)). When the $HG_{n0}$ beam is obstructed until only one pair of faculae at the outermost edge is retained, the reconstructed light intensity distribution is just the result of the interference of faculae on both sides. Because the remaining pair of spots come from the same beam and have good coherent characteristics. The interference between them can be explained by the classical Young's double-slit interference principle. Assume that the interference light path satisfies Fresnel approximation, the intensity distribution of Young's fringes on the observation screen can be expressed as $I(x) = 4I_0 cos^2\left(\pi\frac{l}{\lambda d}x\right)$, and the coordinate of the $m$-th order bright line is $x = \frac{\lambda d}{l}m$. $d$ is the transmission distance from the obstacle plane to the observation plane, $x$ is the transverse coordinate axis with the beam center as the origin, and $l$ is the distance between the double 'slits'. Here, the 'slits' can be approximately regarded as the distance between the peak intensity locations of the two outer edged spots, which is determined by the corresponding order of the HG beam. And $l = 2\alpha\sqrt{2n+1}\sqrt{1+(a/z_R)^2}\omega_0$, $a$ is the distance from the beam source plane to the obstacle plane, $z_R$ is the Rayleigh length, and $\alpha$ is the proportional coefficient between the peak distance of the outer edge spots and the beamwidth. Assume the outer edges of the beam are not affected by occlusion, the far-field divergence half-angle $\theta = \sqrt{2n+1}\frac{\lambda}{\pi\omega_0}$ remains unchanged, which limits the width $w$ of the far-field beam after transmission, and there is $\omega = 2\sqrt{2n+1}\sqrt{1+((a+d)/z_R)^2}\omega_0$. If $n$ is even, the initial phase difference of the retained outer edge spots is zero. Within this width range, $x_{n/2} = \frac{\pi n d}{4\alpha z_R\sqrt{2n+1}\sqrt{1+(a/z_R)^2}}\omega_0$, only 0, ±1..., ± n/2 level bright spots can be formed by constructive interference. If $n$ is odd, the initial phase difference of the retained outer edge spots is π. The center of the reconstructed light field is a dark line due to the destructive interference, and the constructive interference forms ±1..., ± (n + 1)/2 bright spots within the beamwidth. Then one can notice that the reconstructed beam by the interference effect in far-field always has a structure similar to that of the original mode $HG_{n0}$. Further consider that in practice, the width of retained outer edge spots is greater than the ideal slit width in Young's interference, which can be regarded as the result of double beam interference modulated by single-slit diffraction, and the far-field light intensity presents a weakening distribution from center to edge under the limited width (as shown in subgraph (s1) of Fig.2 (a) and Fig.2 (b)). What's more, if the width of the obstacle continues to increase to the peak value of the outer edge spots, it can be considered that the distance between the double slits increases, the bright pattern interval decreases, and more levels of spots will be derived in the far-field (as shown in subgraph (s2) of Fig.2 (a) and Fig.2 (c)). The experimental results of the near-field and far-field light intensity distributions of $HG_{20}$ partially obstructed by the edge spots both prove the interference phenomenon (as shown in Fig.2 (f2) and Fig.2 (f3)).

When the high order HG beam is obstructed from one side in turn, it can be seen from Fig.2 that the obstructed beam cannot restore its original structure, and the far-field light intensity is the diffraction distribution of the remaining beam. This distribution is very close to the corresponding low order HG beam intensity distribution, resulting in the decrease of similarity S (as shown in subgraph (s3) and (s4) of Fig.2 (a)). This is because the unobstructed end maintains the original far-field divergence angle, and the other end expands to the shadow area of the obstacle due to diffraction. The energy flows to the defect, and finally presents a stable low order HG beam intensity distribution (as shown in Fig.2 (d) and Fig.2 (e)). Because the original distance between the remaining spots is maintained, it will not interfere to form new light spots.

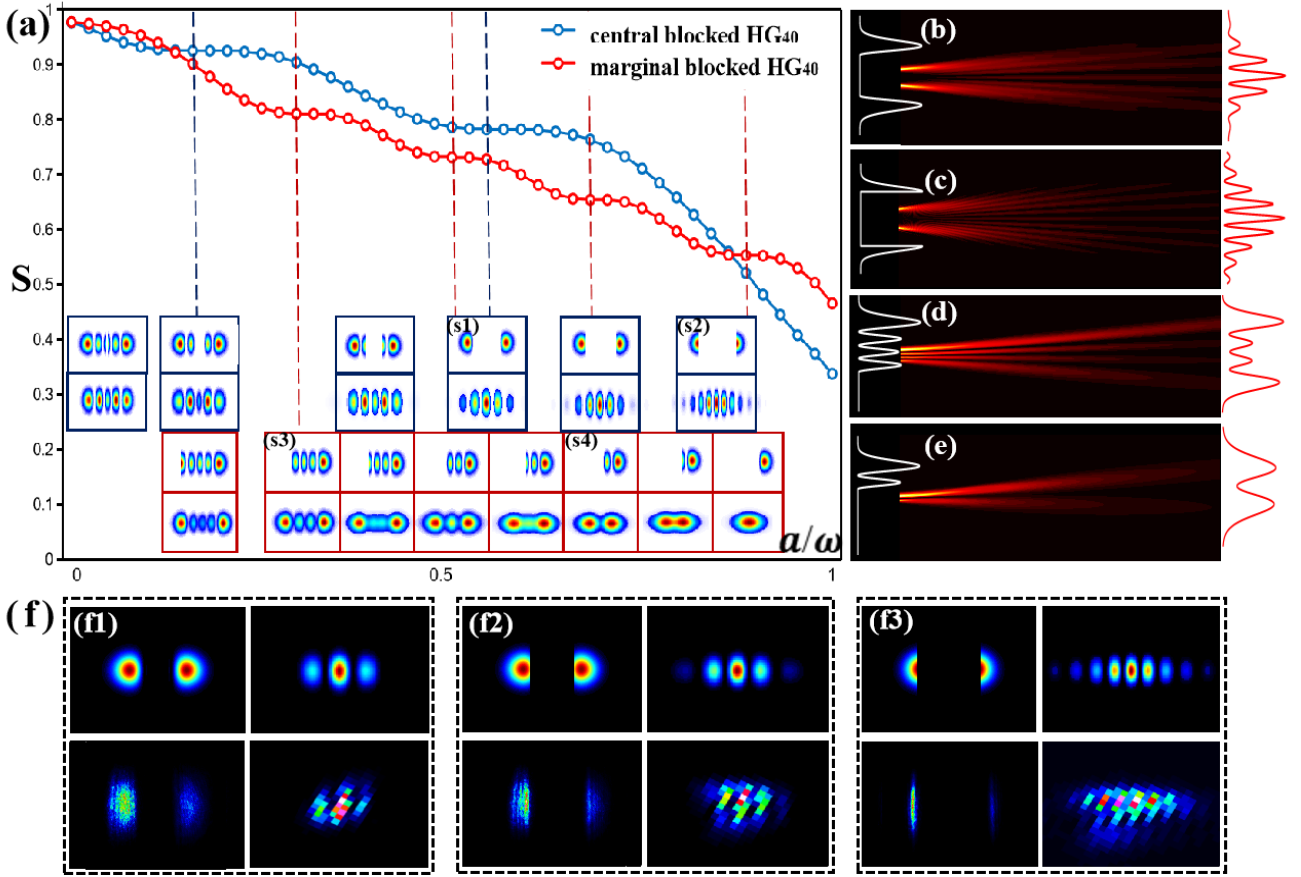

**Fig. 2.** The similarity change curve between the far-field reconstructed beam of $HG_{40}$ occluded by obstacles with different widths from the center or the edge and the original beam. The light intensity distribution of the reconstructed $HG_{40}$ beam obstructed to varying degrees on the observation plane are shown in subgraph (s1) - (s4) of figure (a). The light intensity variation process of the obstructed $HG_{40}$ beam corresponding to subgraph (s1) - (s4) of figure (a) in the x-z plane is shown in figure (b)-(e). And the intensity distributions on the x-axis before and after transmission are also shown by the white and red curves. The experimental results with the interference phenomenon of the near-field and far-field light intensity distribution of $HG_{20}$ partially obstructed by the edge spots are shown in figure (f).

According to the analysis results of the one-dimensional HG beam, its principle can be extended to a two-dimensional high-order HG beam. Since any light field can be decomposed according to any selected orthogonal basis, $U(x_\perp, z) = \sum_n c_n \phi_n(x_\perp, z)$, and the modal content is propagation unchanged, the self-healing degree of the beam can be effectively measured by the modal spectrum of the obstructed beam [21], and the modal transformation in some obstructed cases can be visually displayed. Let's take the $HG_{32}$ mode as an example. When the $HG_{32}$ beam is obstructed to retain a circle of peripheral light spots, the central light spot is reconstructed due to interference of all the edged spots. At the same time, the light spot intensity of the surrounding vertex angles decreases, and the light spot intensity of the surrounding edges increases (as shown in Fig.3 (a)). When the $HG_{32}$ beam is obstructed to only retain the peripheral vertex spots, the pair of spots at a certain interval interfere with each other, the spot array will be reconstructed, and the light intensity is further concentrated to the center (as shown in Fig.3 (b)). When the $HG_{32}$ beam is obstructed to retain only one row and one column of spots, the array can be reconstructed by the constructive interference between the remaining spots in the shadow area within the width of the existing spots. The interference between the two columns of spots distributed at right angles will make the light spots inclined along the diagonal direction (as shown in Fig.3 (c)). In addition, through simulation calculation, it can be found that when the high order HG beam is obstructed to retain only three vertex spots, the spots still retain the original width range in x and y directions and the far-field divergence angles in two directions. Through the interference between the only three vertex spots, the structure information of the original beam can be fairly well reconstructed in the far-field. When one or both edges of the $HG_{32}$ beam are obstructed, the remaining spot continues to propagate to the far-field in the form of diffraction, which can retain the unobstructed spot structure information. At the same time, the energy flows in the transverse section from the unobstructed part to the obstructed part, resulting in the enhancement of the light intensity near the obstructed side in the far-field. The far-field reconstruction distribution of the beam with edge obstruction along the nodal line is very close to the lower order HG beam. For example, due to edge obstruction, the $HG_{32}$ beam becomes the lower order $HG_{31}$, $HG_{22}$ and $HG_{21}$ beams after reconstructing in the far-field (as shown in Fig.3 (d), (e) and (f)).

Summarizing the above simulation analysis, we find that during the transmission of the obstructed beam, while the energy flows laterally due to diffraction, the remaining spots with a certain interval can generate the original number of spots within the existing width due to mutual interference, so as to maintain the corresponding structural information, that is, the number of nodes and main modal components. However, due to the different positions of obstruction, the interference will lead to the concentration of light intensity to the central spots and the inclination of the spots to the defect. Due to the different areas of obstruction, the proportion of main modal components will decrease to varying degrees. When the high order HG beam is obstructed by the outer edge spots of the whole row or column, the beam will expand to the low-order HG distribution of the remaining spots, and the modal components will change accordingly. This is because the divergence angle of the unobstructed side remains unchanged, the number of nodal lines of the beam decreases on the obstructed side, and the spots on the obstructed side gradually expand by diffraction. After the beam energy flows to the equilibrium state, it becomes a low-order HG mode distribution.

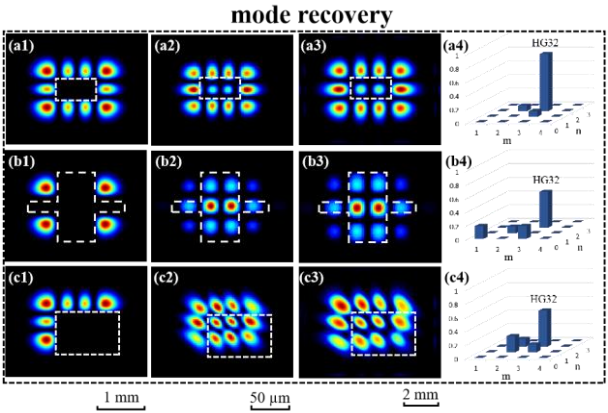

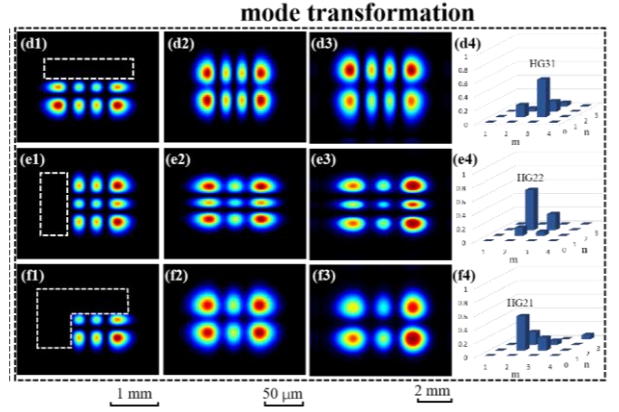

**Fig. 3.** The light intensity distribution of $HG_{32}$ on the observation plane after being occluded by different obstacles. The first column is the intensity distribution of the obstructed $HG_{32}$ on the obstacle plane. The second column is the light intensity distribution of the corresponding obstructed $HG_{32}$ on the observation plane. The third column is the far field intensity distribution of the obstructed $HG_{32}$ mode without lens. The fourth column is the modal spectral components of the reconstructed beam. The white dotted line represents the occluded part of the obstacle. The same goes for the figures below.

In order to show the process of reconstructing the internal spots of the occluded HG beam more clearly, the transverse and longitudinal light intensity changes in the obstructed shadow area are calculated, and the internal energy flow of the beam at a certain transmission distance is analyzed. The energy flow of an optical field is usually indicated by the Poynting vector defined as [41]:

$$\vec{S} = \vec{S_z} + \vec{S_\perp} = \frac{1}{2\eta_0}|E|^2 z + \frac{i}{4\eta_0 k}\left(E\nabla_\perp E^* - E^*\nabla_\perp E\right). \quad (8)$$

Where $\nabla_\perp$ is the curl of the transverse field, $\eta_0 = \sqrt{u_0/\varepsilon_0}$ is the impedance of the free space, and the second term represents the direction of the transverse energy flow. For the case of Fig.3 (b1), there is no light intensity distribution in the shadow area initially, the edge spots appear first, and finally all internal spots appear, and the light intensity concentrates to the center (as shown in Fig.4 (a)). For the case of Fig.3 (c1), the initial axial section is the single spot distribution after the beam is obstructed, and the spots appear in turn with the flow of energy to the shadow area (as shown in Fig.4 (b)). In the process of transmission, the energy flows from the nearby area to the shadow area of obstruction in the beam cross-section, and further forms bright spots due to constructive interference and dark lines due to destructive interference. Finally, the beam forms the corresponding spots distribution within the limit of far-field divergence angle and realizes the redistribution of energy at the same time.

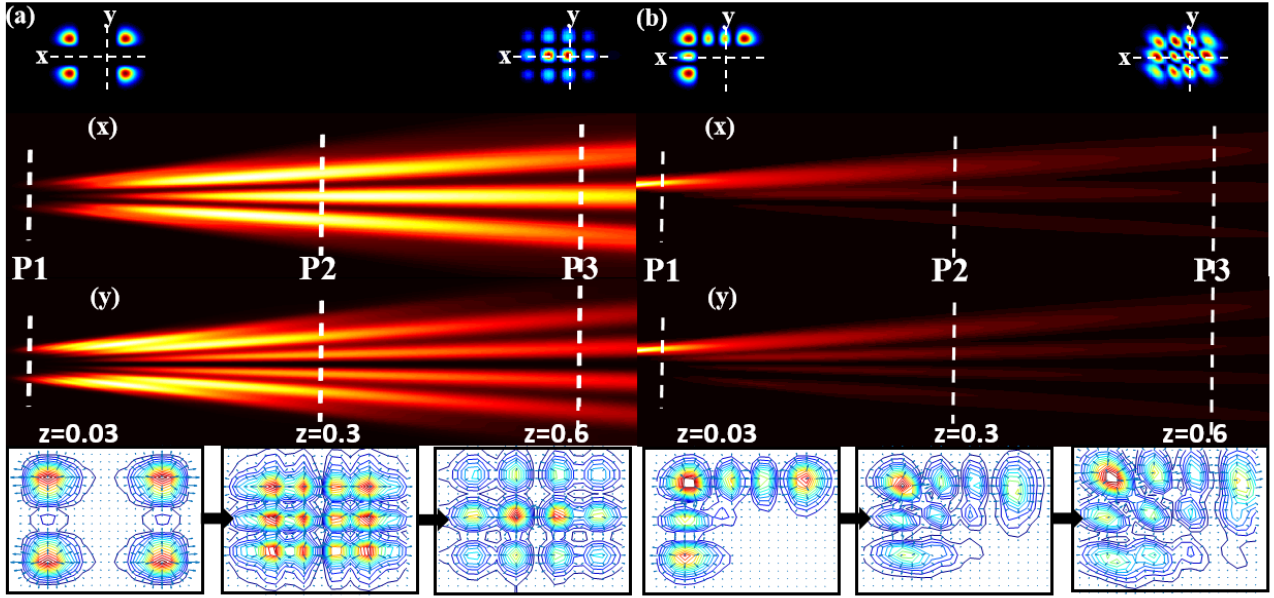

**Fig. 4.** The transverse and longitudinal light intensity changes in the obstructed shadow area and transverse energy flow $\vec{S}_\perp$ of obstructed HG beams at different transmission distances.

HG beam is one kind of propagation-invariant laser beam, and its near-field and far-field shapes are the same [20]. The phase of the unobstructed HG beam is $\varphi(z) = kz + k\frac{(x^2+y^2)}{2R} - (m+n+1)tan^{-1}(\frac{\lambda z}{\pi \omega_0^2})$. When it propagates to a certain distance ($z > 5z_R$), there is $tan^{-1}(\frac{\lambda z}{\pi \omega_0^2}) \to \frac{\pi}{2}$ for the par-axial beam. From the second term $k\frac{(x^2+y^2)}{2R}$ of the formula, the field distribution of the beam in the x-y plane is affected by the occlusion of obstacles on the item of $(x^2 + y^2)$. At the beam waist of the obstructed and focused beam after the thin lens, there is $(x^2 + y^2)$, and $k\frac{(x^2+y^2)}{2R} \to 0$, then $\varphi(z)$ changes to a near plane wave, and the effect of occlusion on the transverse distributed phase item is eliminated. At the same time, the beam waist position after focusing has the smallest spot size, the most concentrated energy distribution, and the best recovery state of the light intensity distribution. On the other hand, when the obstructed beam propagates to a long distance without a focusing lens, its par-axial wavefront flattens gradually, also to obtain $(x^2 + y^2)$, $k\frac{(x^2+y^2)}{2R} \to 0$, and $\varphi(z)$ mainly depends on the first term kz. The influence of obstructions on the beam is weakened, and the light field distribution is gradually reconstructed. Therefore, the obstructed beam can be reconstructed in the far-field or at the beam waist position after passing through the focusing lens.

In addition to the HG beam in the Cartesian coordinate system, the LG beam in the polar coordinate system belongs to propagation-invariant laser beams and is also robust to the occlusion of obstacles. Different from the axial symmetry of the HG beam, the LG beam has circular symmetry, and the occlusion of obstacles with different shapes has different effects on it. Previous studies believe that the self-healing process of LG beam requires high spatial frequency components of the light field, that is, when modulated by a low spatial frequency filter (opaque disk), LG beam can realize self-healing [22]. The outer ring of the LG beam can be regarded as composed of countless pairs of spots symmetrical about the center of the circle. Each pair of spots with an interval of the diameter of the outer ring interfere with each other in transmission to form spots with the same bright and dark spacing. The bright spots are symmetrical about the center of the circle at a certain distance and coupled into an internal bright ring (as shown in Fig.5 (a)). Because the outer edge of the beam is not affected by obstacles, the beam divergence angle remains unchanged, which limits the far-field beam range. Therefore, the interference fringes will not expand infinitely, but only generate the corresponding number of fringes internally. When the radius of the opaque disk is further expanded, the spacing between the pairs of outer edge spots increases, the number of interference fringes increases and the spacing decreases within the same beam range, resulting in the distortion of the obstructed LG beam (as shown in Fig.5 (b)). When the outer ring is obstructed into four equally spaced parts, part of the internal light spots generated by interference are missing, which are not enough to couple into a ring, thus forming a multi-layer square bright spots lattice (as shown in Fig.5 (c)). When the outer ring is blocked into three equally spaced parts, the internal light spots generated by interference are further lost, thus forming a multi-layer triangular bright spots lattice (as shown in Fig.5 (d)). Compared with the HG beam, the partial obstruction of the external light field is easier to destroy the circular symmetry of the LG beam, which makes the internal spots reconstructed by the interference quite different from the original structure, and makes the self-healing effect of the beam worse.

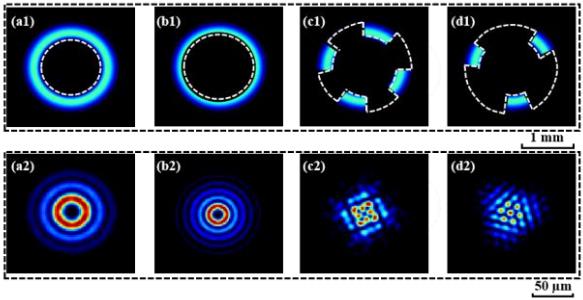

**Fig. 5.** The light intensity distribution of $LG_{23}$ on the observation plane after being occluded by different obstacles. The first line is the intensity distribution of the obstructed $LG_{23}$ mode on the obstacle plane. The second line is the obtained beam intensity distribution of the corresponding obstructed $LG_{23}$ mode on the observation plane.

*2.2 Self-healing and transformation of coherent and incoherent superposition of two modes*

The superposition of two HG modes with the same orders, waist position, and waist parameters can produce interesting optical composed structures, and also has self-healing ability after obstructions. The superposition of the HG modes can be coherent or incoherent. In the case of incoherent superposition, the light intensity distribution on the observation plane is the direct superposition of that of each mode transmitted to the observation plane separately. To coherent superposition, the phase and frequency of each mode need to be precisely locked. The field on the source plane is the superposition of complex amplitude of light field as Eq. (2), and the two modes still maintain a coherent state after the occlusion. Then the intensity distribution obtained on the observation plane is also a coherent one, whether the order of the modes gets changed.

The obstructed two-mode-superposition beam has a similar property on self-healing and transformation as the single-mode beam. If the central part of the beam is obstructed, without changing the beamwidth of each component mode, the original beam structure can be reconstructed on the observation plane for both coherent and incoherent conditions. If the edge of the beam is obstructed, the obtained pattern on the observation plane will change for both coherent and incoherent conditions. And the obtained changed beam will be the coherent or incoherent superposition of the varied component modes, which leads to quite different far-field patterns. As shown in Fig.6 (a), $HG_{03}$ and $HG_{30}$ are incoherently or coherently superimposed to both form a four-petal flower structure. For the coherent superposition, there is a $\pi/2$ phase difference between the two HG modes. If the position of the central stamen is obstructed, the beam can return to its original structure in the far-field for both two states. However, the central intensity of the reconstructed beam is stronger than that of the original beam. If the position of one edged (take the upper one for example) petal is obstructed, the beams formed by incoherent and coherent superposition have quite different distortions. The obtained changed structures will correspond to the patterns formed by the coherent and incoherent superposition of $HG_{02}$ and $HG_{30}$, respectively. Similarly, $HG_{40}$ and $HG_{22}$ modes can superimpose to form a symmetric structure of three lobes in coherent or incoherent condition, which both can recover to original patterns with stronger central intensity in the far-field when the intermediate spot is obstructed. When occluding one edge section of the beam, the altered patterns are the coherent and incoherent superposition of $HG_{20}$ and $HG_{12}$ modes. By comparing the obtained far-field patterns with the original one, it can be seen that incoherent superposition condition results in less beam distortion on the beam structures.

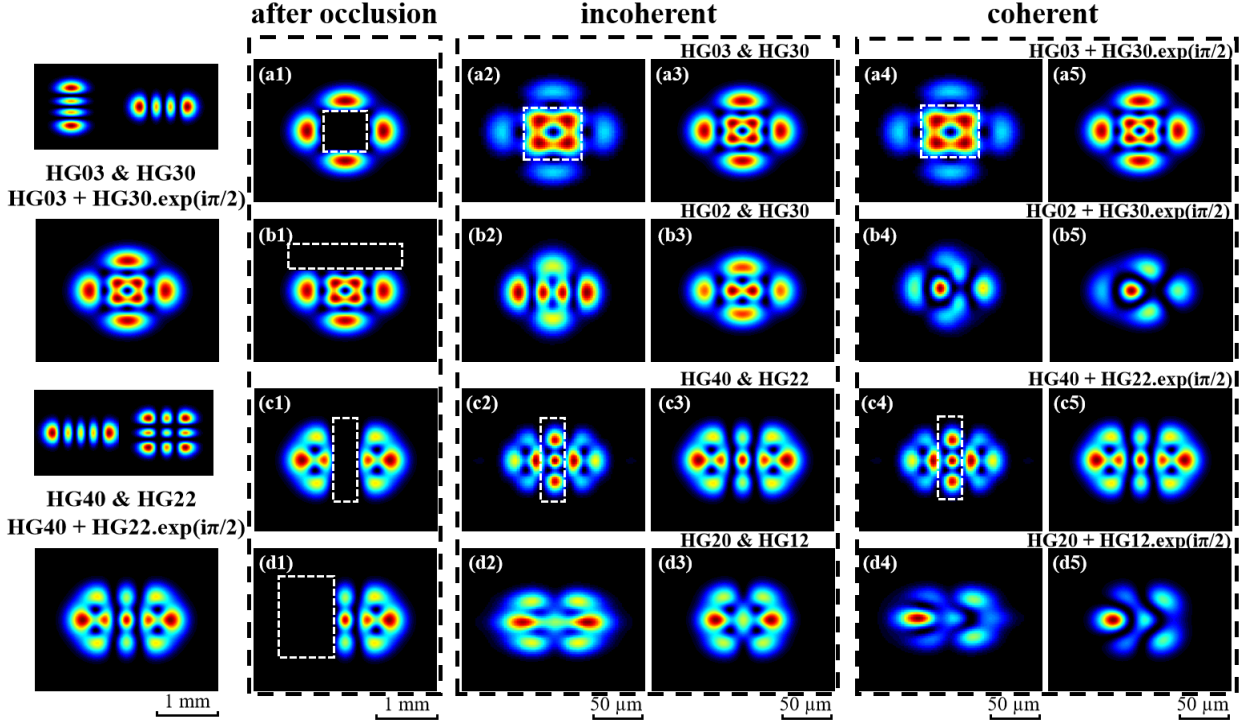

**Fig. 6.** The light intensity distribution on the observation plane of the structured beam formed by incoherent and coherent superposition of two HG modes obstructed by different position obstacles. The first column is $HG_{03}$ & $HG_{30}$ (or $HG_{03} + HG_{30}.exp(i\pi/2)$) and $HG_{40}$ & $HG_{22}$ (or $HG_{40} + HG_{22}.exp(i\pi/2)$) light intensity distribution without occlusion, the beam patterns can be obtained in both coherent and incoherent states. The second column is the light intensity distribution of the obstructed optical composed structure on the obstacle plane. The third and fifth columns are the light intensity distribution of the obstructed optical composed structure on the observation plane. The fourth and sixth columns are incoherent and coherent superposition of single HG mode without occlusion, to make comparison with the third and fifth columns.

## 2.2. Self-healing characteristics of incoherent superposition of multiple modes

From the above section, it can be seen that the incoherent superposition of modes has a better recovery result for the structured beams in the edge occlusion condition. The optical structures of circular symmetry can be constructed by the incoherent superposition of multiple HG modes. The circular flat-topped beam can be formed by the incoherent superposition of all modes of adjacent orders m+n in a certain proportion. For example, the incoherent superposition of all modes with order m+n=0,1,2 ($HG_{00}$&$HG_{01}$&$HG_{10}$&$HG_{02}$&$HG_{20}$&$HG_{11}$) can form a circular flattop structure (as shown in Fig.7 (a1)). The concentric ring structure with (m+n+1)/2 layers can be formed by the incoherent superposition of all modes of odd order m+n in the same proportion. For example, the incoherent superposition of all modes with order m+n=3 ($HG_{03}$&$HG_{30}$&$HG_{12}$&$HG_{21}$) can form a double-ring structure (as shown in Fig.7 (c1)). The incoherent synthesis of flat-topped structures with larger areas and ring structures with more layers can be analogized. Super-Gaussian beam and LG beam have circular flat-topped and multi-layer ring structures respectively. The particularity of structure shape makes them have strong application value, but it is difficult to reconstruct their original structure after encountering the interference of obstacles. The incoherent superposition of multiple HG modes can generate beams with similar structures and has good self-healing ability for the occlusion of obstacles.

This kind of circular symmetric synthetic optical structure, super-Gaussian beam and LG beam with similar shapes are numerically analyzed after being occluded by different obstacles. When the circular flat-topped structure composed of $\sum HG_{mn}$ (m+n=0,1,2) incoherent superposition is obstructed by a 45° fan or a rectangular obstacle with the side length of $6w_0$, the circular light spot can recover without damage, and the flat-topped characteristics are well remained good on the observation plane (as shown in Fig.7 (a3) and (a5)). For the third-order super-Gaussian beam with a similar structure, the flat-topped characteristic of the beam is destroyed after being obstructed by the same sector and evolves to a Gaussian-like distribution, as shown in Fig.7 (b3). And when it is obstructed by the same rectangle above, the light intensity further concentrates to the center and generates an additional diffraction ring, as shown in Fig.7 (b5). When the double-ring structure composed of $\sum HG_{mn}$ (m+n=3) incoherent superposition is covered by a 45° fan or a rectangular obstacle with the side length of $6w_0$, the circular light spot is recovered without damage, and the double-ring structure is basically clear on the observation plane (as shown in Fig.7 (c3) (c5)). For the $LG_{11}$ beam with a similar structure, the ring structure is obviously distorted after being obstructed by the same sector, as shown in Fig.7 (d3). After being obstructed by the same rectangle, the beam recovers well, which is consistent with the previous analysis as shown in Fig.7 (d5). From these simulation results we can conclude that, for the HG modes composed synthetic optical structures, whether it is a fan-shaped obstacle with the radial occlusion or a rectangular obstacle with the central occlusion, after a certain distance of transmission, the light intensity compensation and the overall structure recovery at the damaged position can be observed. While, we can notice that, the reconstructed light intensity slightly decreases in the direction of occlusion for the radial occlusion, and the reconstructed light intensity is slightly concentrated towards the center for the central occlusion. However, super-Gaussian beams will lose their flat-topped characteristics after being occluded by obstacles with different shapes, and the light intensity will concentrate on the center to a large extent. LG beam has a strong anti-interference ability only when it is obstructed by the center, and distortion will occur when it is obstructed by the edge, resulting in the loss of circular symmetry.

The good anti-occlusion performance of this optical structure synthesized by HG modes is mainly due to three reasons. Firstly, it is composed of an incoherent superposition of multiple HG modes. When one mode is distorted by the occlusion, the incoherent superposition of light intensity of other modes can weaken this effect. This is similar to the well-known reality of an incoherent beam is more robust than a coherent beam. Secondly, based on the self-healing property for central occlusions of HG modes, the incoherent superposition formed flat-topped beam can well reconstruct, which is different from the single beam in other modes. And thirdly, the circular symmetry has good structural stability which is also contributed by the above two aspects.

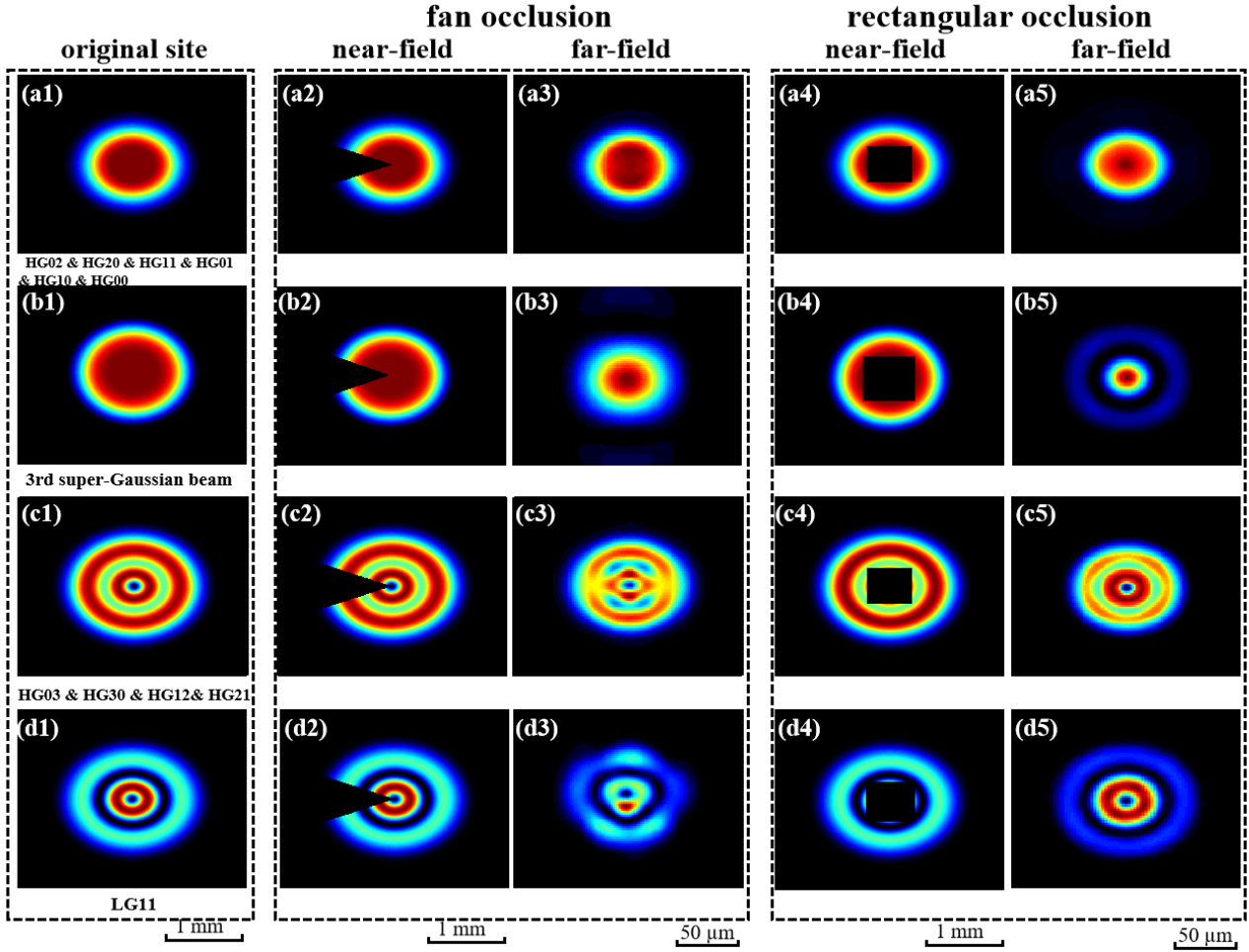

**Fig. 7.** The self-healing effect of obstructed circularly symmetric synthetic optical structure formed by incoherent superposition of multiple HG modes, super-Gaussian beam, and LG beam. The first column is the intensity distributions of the unobstructed optical structures. The second and fourth columns are the light intensity distributions of the optical structures with radial sector occlusion and central rectangle occlusion on the obstacle plane. The third and fifth columns are the light intensity distributions of the corresponding obstructed optical structures on the observation plane.

### 3. Experiments and discussions

The self-healing and transformation characteristics of structured beams are investigated experimentally with the setup shown in Fig.8. The HG or LG modes composed structured beams can be directly generated in the laser oscillator with a large pumping Fresnel number [42-44] or synthesized by a spatial light modulator (SLM) loaded with holograms method [45,46]. Based on the principle of selective pumping, gain distribution and high non-linearity, a specially designed microchip laser (0.5 mm thick Nd: YAG and 1 mm thick LiTaO3 stacked together) can be used to generate complex structured beams directly. The intensity distribution of the laser beam can be easily adjusted by changing the pump parameters, such as the area of the pump region, the incident angle, and the power. The wavelength of the output structured laser beam is 1.064 µm and the waist radius of the basic mode is 53 µm with a divergence angle of about 6.3 mrad. When the beam reaches z=0.5m, it is obstructed by opaque obstacles, and the obstructed beam is focused by a thin convex lens with a focal length of 0.1 m. After passing through the beam splitter, one beam is received by a photodetector to adjust the coherent or incoherent states of the beams, and the beat frequency spectrum can be observed for incoherent superposition states and disappears in a coherent superposition state. Another beam enters the charge coupled device (CCD). The distribution of the light field at the minimum spot (the waist position of the beam) is measured by adjusting the z position of CCD properly. Other conditions remain unchanged, and the measurement after removing obstacles is taken as a reference.

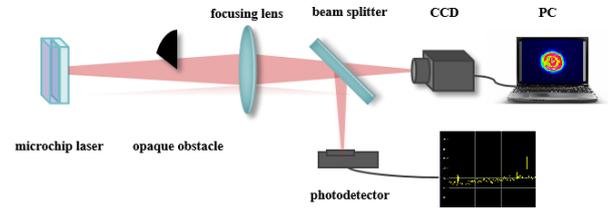

**Fig. 8.** The experimental setup for generating structured beams and observing their self-healing characteristics.

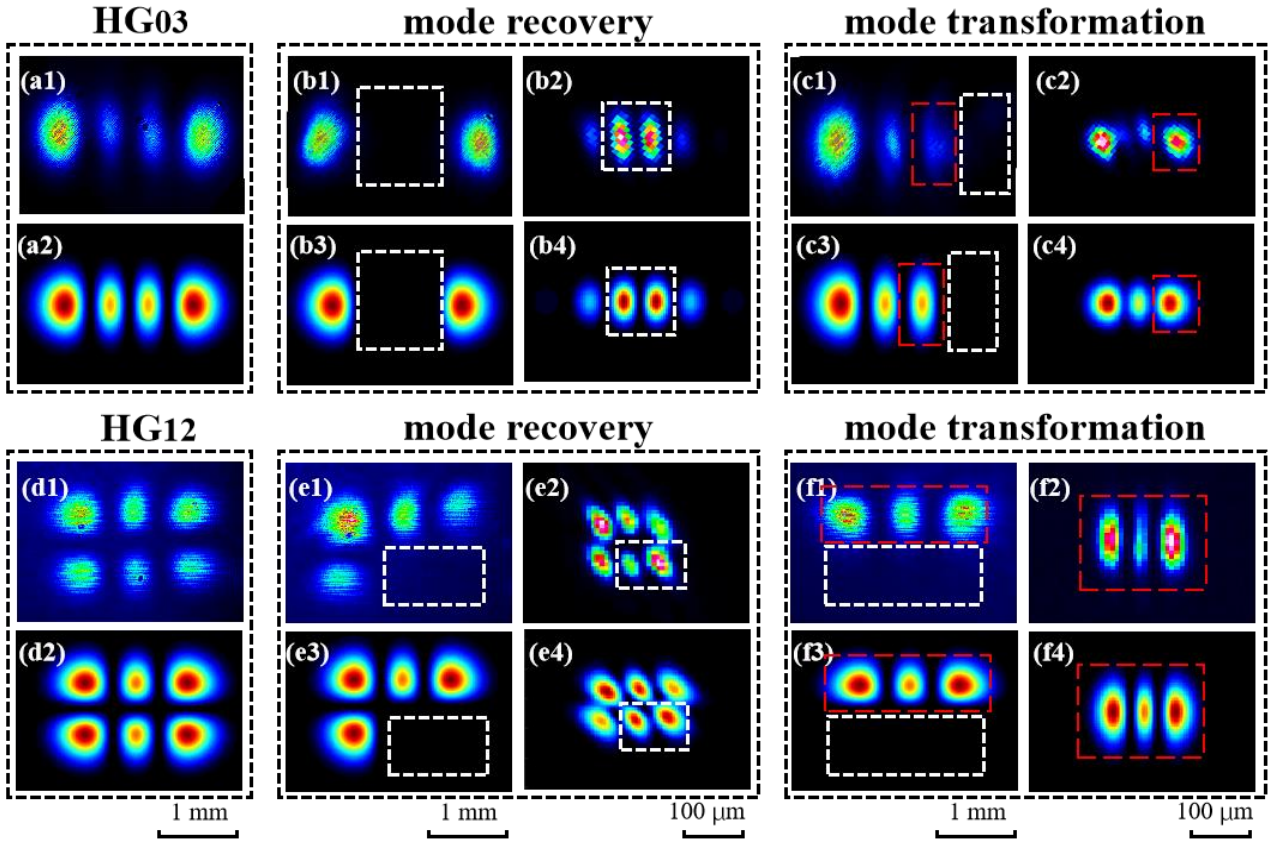

**Fig. 9.** Experimentally measured (row 1 and 3) and corresponding simulated beam patterns (row 2 and 4) of self-healing and transformation cases for single HG beam with different occlusions. The first column is the near-field beam patterns without occlusion, the second and fourth columns are the near-field beam patterns with occlusion, and the third and fifth columns are the far-field intensity distribution of the obstructed beams.

The experimental and simulation results of the far-field intensity distribution of a single HG beam with different occlusions are shown in Fig.9. For example, $HG_{03}$ mode is obstructed at the middle two faculae, all faculae can be recovered in the far-field, and the light intensity is concentrated to the center. If the edged facula is obstructed, the $HG_{03}$ mode will evolve into $HG_{02}$ mode's pattern at beam waist after focusing by a lens. Similarly, if $HG_{12}$ mode is obstructed at the lower right two faculae, all faculae can be recovered in the far-field, and the spots are inclined diagonally.

However, if the whole line of faculae in the lower part is obstructed, it will evolve into $HG_{02}$ mode's pattern at beam waist after focusing by a lens. And more experimental examples can show the same results. These experimental results are consistent quite well with the theoretical analysis and simulations. It's experimentally proved that when the obstructed single HG beam retains the outer edge vertex spots, the original beam mode can be restored after the beam propagates to the far-field. However, due to the different interference between the remaining spots with different distributions, the light intensity distribution has a certain change compared with the original one (as shown in Fig.9 (b) and (e)). While if the entire row or column of outer edge spots are occluded, the distribution of light field will transform to the lower order corresponding to the number of the remaining faculae (as shown in Fig.9 (c) and (f)).

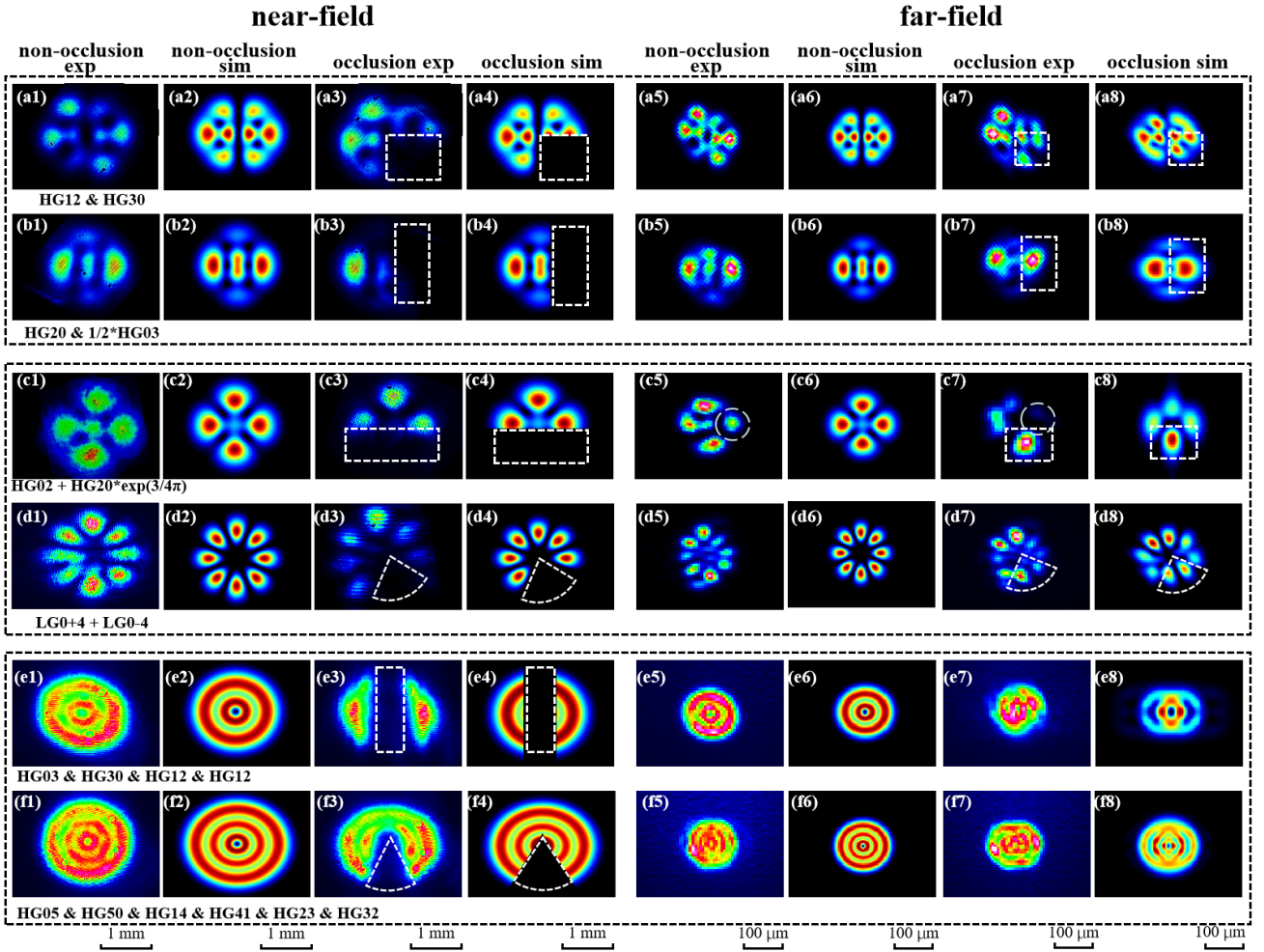

**Fig. 10.** The experimental results of self-healing and transformation of the obstructed structured beams formed by coherent and incoherent superposition of HG or LG modes. Experimental results (column 1,3,5,7) and corresponding simulations (column 2,4,6,8) of near-field and far-field intensity distribution with and without occlusion are presented. Columns 1 and 2 are the near-field intensity distribution without occlusion, columns 5 and 6 are the corresponding far-field intensity distribution.

Columns 3 and 4 are the near-field intensity distribution after occlusion, columns 7 and 8 are the corresponding obstructed far-field intensity distribution.

The experimental results and corresponding simulated results of near-field and far-field intensity distribution of complex optical structured beams formed by the coherent and incoherent superposition of HG and LG modes with and without occlusion are shown in Fig.10. Line (a) shows the HG modes composed structured beam formed by the incoherent superposition of $HG_{12}$ and $HG_{30}$, and its quarter area is obstructed. The light intensity at the defect area can be supplemented, and the original structure is recovered and basically clear as shown in Fig.10 (a7). Line (b) shows the structured beam formed by the incoherent superposition of $HG_{20}$ and $0.5HG_{03}$, and its 1/3 area on the right side is obstructed. The intensity of light recovered to a symmetrical distribution. However, the beam structure evolved to the incoherent superposition of $HG_{20}$ and $0.5HG_{02}$. Line (c) shows the structured beam formed by $HG_{02}$ and $HG_{20}*\exp(i3/4\pi)$ coherent superposition. In the far-field, the beam pattern rearranges and produces a certain degree of distortion compared with the original structure. This is induced by that the outer edge spots of the whole column on the right side of mode $HG_{02}$ is blocked by obstacles and evolves to mode $HG_{01}$. The obtained unsymmetric pattern distribution reveals the coherent superposition of the changed HG modes. Note that what in the white circle labels is a relatively weak spot due to uneven initial light intensity. In the case of incoherent superposition, the light intensity is superimposed directly, and the deformation is weakened due to the complementary effect. For the coherent combination, both amplitude and phase of the modes are coupled with each other, and the influence of this deformation is enhanced, which leads to the larger distortion of the HG modes composed structured beam. When the beam is obstructed by all the outer edge spots, the mode order is reduced, and the incoherent combination can still get the symmetrical distribution of similar structures, while the coherent combination will get quite different interference structures. Compared with the coherent superposition of HG modes in line (c), the petal structure formed by the coherent superposition of $LG_{04}$ and $LG_{0-4}$ modes in line (d) has better recovery after the occlusion. This is because the LG mode of annular light intensity has less loss of outer edge information under the occlusion of a small-angle fan, so it can be restored to the original mode. In addition, the distortion of LG modes with opposite topological charges after reconstruction can complement each other. Line (e) and line (f) show the multi-ring structured beam formed by the incoherent superposition of multiple HG modes. After being obstructed by rectangular or fan-shaped obstacles, the circular pattern is hardly damaged, the inner ring structure is basically clear, and the original beam pattern structure is well reconstructed. It can be seen that the incoherent superposition of multiple HG modes further weakens the influence of occlusions on the light field distribution, and improves the anti-occlusion characteristic of HG modes composed structured beams.

From the experimental results, the structured beam formed by the incoherent superposition of HG modes has the strong ability of self-healing to occlusion, and its intrinsic mechanism is induced to the self-recovery property of single HG mode and the complementary effect of the incoherent superposition of multiple HG modes.

## 4. Conclusions

The self-healing and transformation characteristics of structured beams formed by incoherent and coherent superposition of HG modes occluded by obstacles are studied both theoretically and experimentally. Firstly, the self-healing and transformation property of the single obstructed HG mode is studied. When the obstructed beam retains more than three vertex spots, the beam can restore its original mode in the far-field due to the interference between the remaining spots. When the obstructed beam loses the whole row and column of spots, the beam will transform to the low-order mode of the remaining spots. Then, the self-healing and transformation properties of structured beams formed by incoherent or coherent superposition of two modes are studied. The self-healing and transformation rules are based on the single-mode condition. Finally, the self-healing characteristics of circularly symmetric optical structures formed by the incoherent superposition of multiple HG modes are studied. It is found that these beams have good self-healing characteristics for radial and central occlusions, and have stronger anti-occlusion ability than coherent superposition cases. And it has better recovery performance when there are more mode components due to the complementary of multiple modes. Experimental results verify the above theoretical analysis and simulations perfectly. The partially coherent optical field of incoherent superposition of multiple modes can be regarded as a random optical field. Previous researches on the self-healing properties of random optical field usually analyze its second-order correlation function, but it is difficult to get an accurate analytical solution for complex optical field. However, our method of source mode incoherent superposition explains the self-healing properties of the random optical field from a new intuitive perspective and explains the self-healing of a single determinate optical field on a more essential physical level. These conclusions are expected to be applied to laser communication, atom optical capture, and optical imaging areas.

**Declaration of Competing Interest**

The authors declare that they have no known competing financial interests or personal relationships that could have appeared to influence the work reported in this paper.

**Acknowledgements**

This research is funded by the National Natural Science Foundation of China (NSFC), grant number 61805013.